 \documentclass[aps,showpacs,prb,groupedaddress,twocolumn]{revtex4}  

\usepackage{graphicx}  
\usepackage{dcolumn}   
\usepackage{bm}        
\usepackage{amssymb}   
\usepackage[english]{babel}

\begin{document}

\title{Surface-induced charge state conversion of nitrogen-vacancy defects in nanodiamonds}
\date{\today}
\author{L.~Rondin$^{1}$}
\author{G.~Dantelle$^{1}$$^{,2}$}
\email{geraldine.dantelle@polytechnique.edu}
\author{A.~Slablab$^{1}$}
\author{F.~Grosshans$^{1}$}
\author{F.~Treussart$^{1}$}
\author{P.~Bergonzo$^{3}$}
\author{S.~Perruchas$^{2}$}
\author{T.~Gacoin$^{2}$}
\author{M.~Chaigneau$^{4}$}
\author{H.-C.~Chang$^{5}$}
\author{V.~Jacques$^{1}$}
\email{vjacques@lpqm.ens-cachan.fr}
\author{J.-F.~Roch$^{1}$}

\affiliation{$^{1}$Laboratoire de Photonique Quantique et Mol\'eculaire, ENS Cachan, UMR CNRS 8537, F-94235 Cachan, France}
\affiliation{$^{2}$Laboratoire de Physique de la Mati\`ere Condens\'ee, Ecole Polytechnique, UMR CNRS 7643, F-91128 Palaiseau, France}
\affiliation{$^{3}$CEA, LIST, Diamond Sensors Laboratory, F-91191 Gif-Sur-Yvette, France}
\affiliation{$^{4}$Laboratoire de Physique des Interfaces et Couches Minces, Ecole Polytechnique, UMR CNRS 7647, F-91128 Palaiseau, France}
\affiliation{$^{5}$Institute of Atomic and Molecular Sciences, Academia Sinica, Taipei 106, Taiwan}

\begin{abstract}
We present a study of the charge state conversion of single nitrogen-vacancy (NV) defects hosted in nanodiamonds (NDs). We first show that the proportion of negatively-charged NV$^{-}$ defects, with respect to its neutral counterpart NV$^{0}$, decreases with the size of the ND. We then propose a simple model based on a layer of electron traps located at the ND surface which is in good agreement with the recorded statistics. By using thermal oxidation to remove the shell of amorphous carbon around the NDs, we demonstrate a significant increase of the proportion of NV$^{-}$ defects in 10-nm NDs. These results are invaluable for further understanding, control and use of the unique properties of negatively-charged NV defects in diamond.
\end{abstract}

\maketitle 

\indent The negatively-charged nitrogen-vacancy (NV$^{-}$) defect in diamond has received considerable interest over the last decade owing to its perfect photostability and to its ground-state electron spin properties, which combine a long coherence time~\cite{Gopi_NatMat2009} and the ability to undergo spin-sensitive optical transitions under ambient conditions~\cite{Jelezko_PRL2004,Manson_PRB2006}. Such properties are at the heart of a wide range of emerging technologies, from imaging in life science~\cite{Chang_NatNano2008,Falkaris_ACSnano2009,Chang_NanoMedecine2009}, to high-resolution magnetic sensing~\cite{Gopi_Nature2008,Maze_Nature2008} and quantum information processing~\cite{Hanson_Science2008,Lukin_Science2009,Neumann_NatPhys2010}. For most of these applications, it is essential to use NV$^{-}$ defects as close as possible to the diamond surface, either to enhance the sensitivity of diamond-based magnetic sensing or to efficiently couple the NV$^{-}$ defect photoluminescence to a photonic waveguide or a microcavity~\cite{Benson_OptLett2009,Santori_APL2009}. This condition can be fulfilled by considering NV$^{-}$ centers hosted in a nanodiamond (ND)~\cite{Chang_NatNano2008,Tisler_ACSNano2009,Rabeau_NatNano2010} or created by shallow implantation in a single-crystal bulk diamond~\cite{Santori_PRB2007}. 
However, it was recently reported that the preferred charged state of NV defects close to the diamond surface is the neutral state~\cite{Santori_PRB2007,Treussart_PhysB2006} NV$^{0}$, which does not exhibit the appealing optical and spin properties of its negatively-charged counterpart. It is then highly desirable to understand the processes leading to charge state conversion, with the aim of producing efficiently NV$^{-}$ color centers close to a diamond surface.\\
\begin{figure}[t] 
\centerline
{\includegraphics[width=8.5cm]{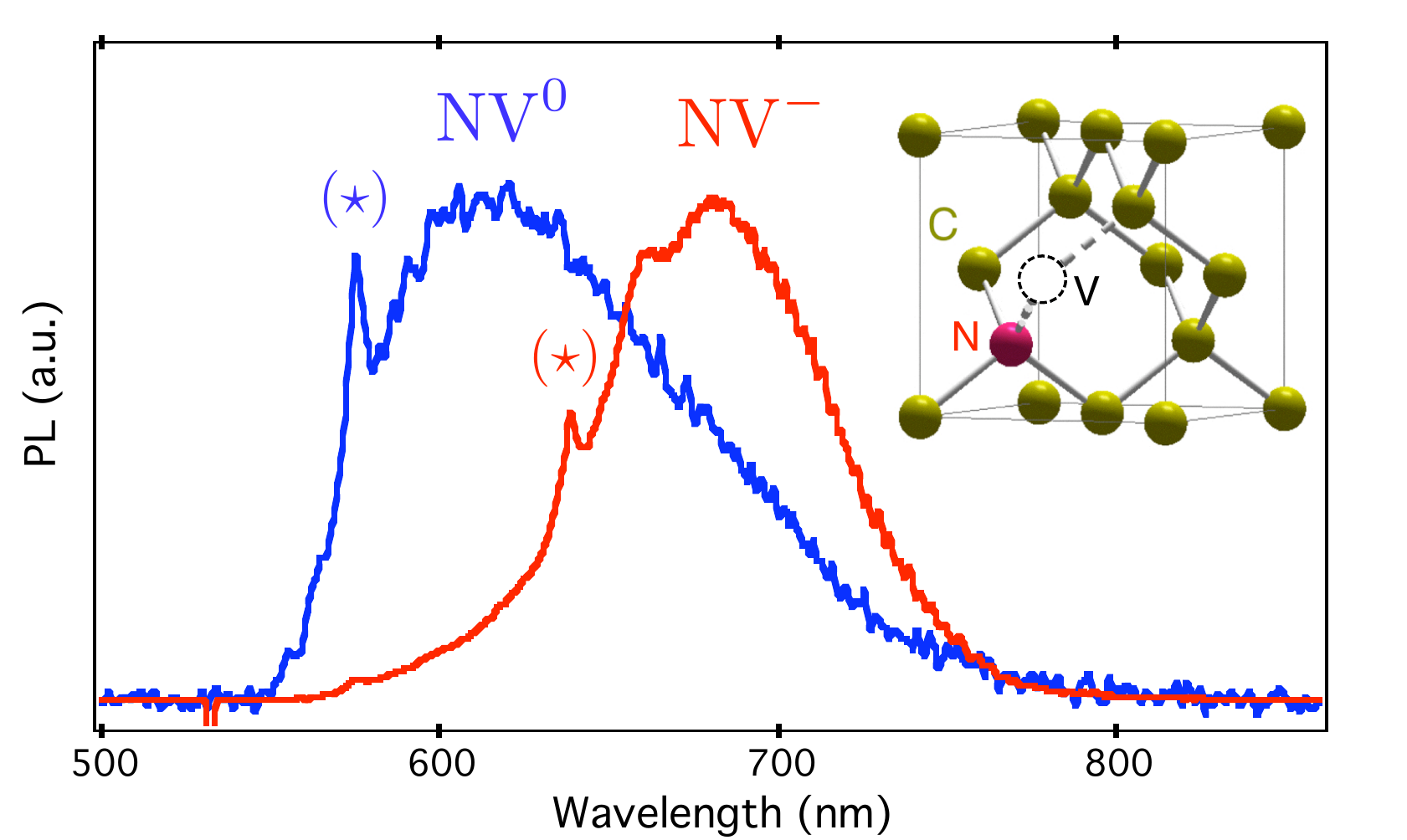}}
\caption{(color online) Photoluminescence (PL) spectra (normalized to their respective maximum value) of single NV$^{-}$ (red curve) and NV$^{0}$ (blue curve) color centers in nanodiamonds. The zero-phonon line ($\star$ symbols) of NV$^{-}$ (resp. NV$^{0}$) emission is located at $637$ nm (resp. $575$ nm). The inset shows the atomic structure of the NV defect, consisting of a substitional nitrogen atom (N) associated to a vacancy (V) in an adjacent lattice site of the diamond crystalline matrix.}
\label{Fig1}
\end{figure}
\indent In this context, we investigate the charge state conversion of single NV defects hosted in NDs of variable size. Since NV$^{0}$ and NV$^{-}$ defects can be discriminated through their photoluminescence spectrum (Fig.~\ref{Fig1}), we determine the proportion $\mathcal{P}_{{\rm NV}^{-}}$ of negatively-charged NV$^{-}$ defects found in an ensemble of NDs containing individual NV color centers, as a function of the ND size. The measurements are done with a confocal microscope combined with an atomic force microscope. We show that $\mathcal{P}_{{\rm NV}^{-}}$ decreases with the ND size, indicating that surface-related effects play a significant role in charge state conversion between NV$^{-}$ and NV$^{0}$. Many studies have indeed established that various impurities at a diamond surface can strongly modify the charge transport properties near the surface \cite{Nebel_Book}. In the specific case of irradiated and annealed diamond, graphitic related defects formed on the diamond surface are possible electron trap candidates~\cite{Ristein_DRM2000} which can ionize donors required for an efficient conversion of the NV defects into the negatively-charged state~\cite{Santori_PRB2007}. Following  Ref.~\cite{Santori_PRB2007}, a simple scheme is then proposed for modeling the experimental observations.  By removing graphitic related defects through thermal oxidation~\cite{Santori_APL2010}, we finally show that the proportion of NV$^{-}$ defects relative to their neutral NV$^{0}$ charge state can be strongly enhanced, especially in the smallest NDs.\\  

We started from commercially available diamond nanocrystals with different particle size distributions (SYP 0.05 and 0.1, Van Moppes SA, Geneva). Such nanocrystals are produced by milling  type-Ib high-pressure high-temperature diamond crystals with a high nitrogen content ($[N]\approx 200$~ppm). The formation of NV defects was carried out using high energy ($13.6$~MeV) electron irradiation~\cite{Saphir_Saclay} followed by annealing at $800^{\circ}$C under vacuum during two hours. The electron irradiation creates vacancies in the diamond matrix while the annealing procedure activates the migration of vacancies to intrinsic nitrogen impurities, leading to NV defect bonding~\cite{Davies,Mita_PRB1996}. After annealing, the irradiated NDs were cleaned using a ``piranha'' solution  which is  a strong oxidizer intended to purify the ND surface~\cite{Piranha}. After sonication and washing with distilled water, aqueous colloidal solutions of dispersed NDs were obtained. A few drops of ammonium hydroxide were then added to the solutions to keep them stable for several weeks. NDs were finally deposited by spin-coating onto a  piranha-cleaned silica coverslip. \\
\indent With the aim of studying the effect of the ND size on the charged state of NV defects, two solutions of irradiated NDs with different size distributions were prepared. These solutions correspond to a respective mean size of $35$~nm (SYP 0.05) and $65$~nm (SYP 0.1), as determined by dynamic light scattering measurements. Due to the broad size distribution, NV defects in NDs with size ranging from $10$ to $100$ nm could be investigated with these two primary solutions. 

\begin{figure}[t] 
\centerline
{\includegraphics[width=8.5cm]{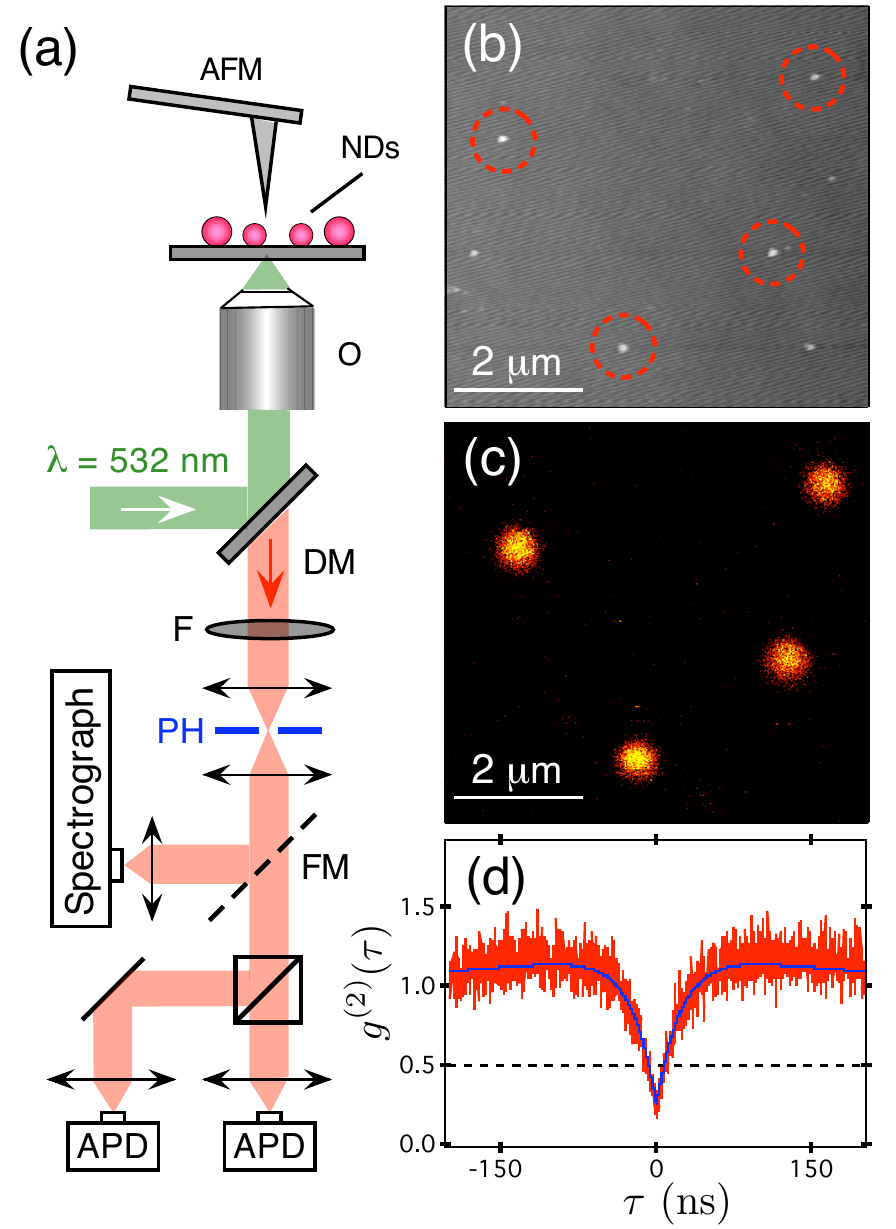}}
\caption{(color online) (a)-Sketch of the experimental setup combining an atomic force microscope (AFM, Asylum Research MFP-3D-BIO) with an inverted confocal microscope. NDs: nanodiamonds; O: oil immersion microscope objective (Olympus, $\times 60$, ${\rm NA}=1.35$); DM: dichroic beamsplitter; F:  $580$~nm long-pass filter; PH: $50 \ \mu$m  diameter pinhole; FM: flip mirror directing the collected photoluminescence either to an imaging spectrograph equipped with a back-illuminated cooled CCD array, or to a Hanbury Brown and Twiss interferometer consisting of two silicon avalanche photodiodes (APD) placed on the output ports of a $50/50$ beamsplitter. (b),(c)-Photoluminescence raster scan of the sample (c) with the corresponding topography image (b). Red dotted circles indicate NDs hosting NV defects. The correspondence between the two images can be easily inferred. (d)-Typical second-order autocorrelation function $g^{(2)}(\tau)$ recorded for a single NV defect in a diamond nanocrystal.}
\label{Fig2}
\end{figure}

\indent The photoluminescence properties of NV defects hosted in NDs were studied using a scanning confocal microscope integrated with an atomic force microscope (AFM), as shown in Fig.~\ref{Fig2}. A laser operating at $532$~nm wavelength was tightly focused onto the sample through a high numerical aperture microscope objective ($\times 60$, ${\rm NA}=1.35$). The photoluminescence emitted by NV defects was collected by the same objective and spectrally filtered from the remaining pump light. Following standard confocal detection scheme, the collected light was then focused onto a $50 \ \mu$m diameter pinhole and directed either to a photon counting detection system for light intensity measurement or to an imaging spectrograph allowing to distinguish between the NV$^{-}$ and NV$^{0}$ emission (see Fig.~\ref{Fig1}). In addition, the AFM was used to record topography images of the sample, thus providing a direct {\it in situ} measurement of the ND size. For each photoluminescent ND, the emission spectrum and the ND size were measured, allowing to correlate the size to the charge state of the NV defects.\\
\indent In order to avoid possible bias in the statistics, the whole study was restricted to NDs hosting single NV defect.  The unicity of the emitter was systematically checked using photon correlation measurements. Since after the emission of a first photon, it takes a finite time for a single emitter to be excited again and then emit a second photon, a dip around zero delay $\tau=0$ appears in the normalized second-order correlation function
\begin{equation}
g^{(2)}(\tau)=\frac{\langle I(t)I(t+\tau)\rangle}{\langle I(t)\rangle^{2}} \ ,
\label{autocorrelation}
\end{equation} 
where $I(t)$ is the photoluminescence intensity at time $t$. The later function is deduced from the histogram of time delays $\tau$ between two consecutively detected photons, recorded by using two avalanche photodiodes placed on the output ports of a beamsplitter, following the standard Hanbury Brown and Twiss scheme (see Fig.~\ref{Fig2}(a)). The observation of an anticorrelation effect at zero time delay, $g^{(2)}(0)< 0.5$, is the signature that a single NV defect is addressed (see Fig.~\ref{Fig2}(d))~\cite{Brouri_OptLett}.

\begin{figure}[t] 
\centerline
{\includegraphics[width=8.5cm]{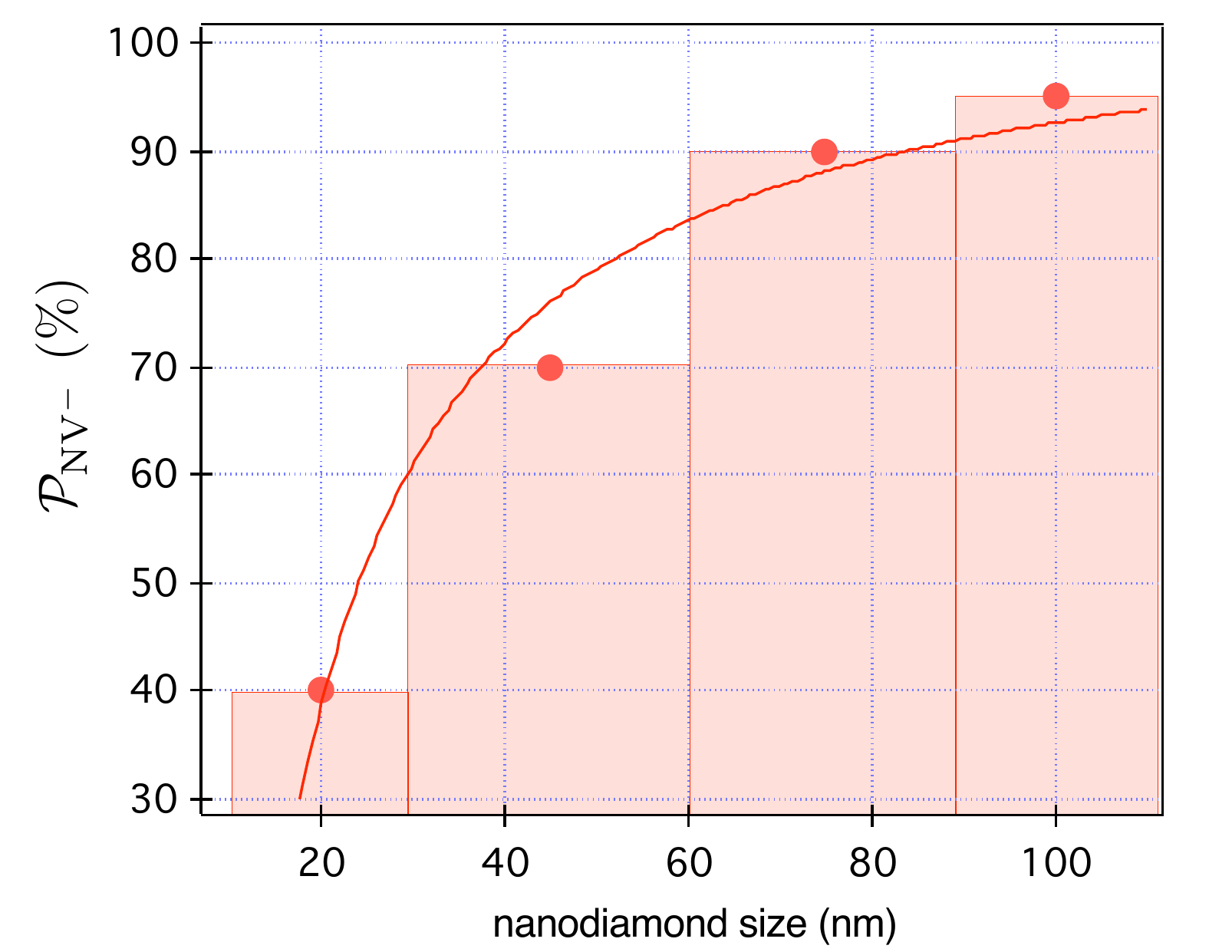}}
\caption{(color online) Proportion of negatively-charged NV$^{-}$ defects $\mathcal{P}_{{\rm NV}^{-}}$ as a function of the size of the photoluminescent ND. The statistics is determined from a set of $146$ single NV defects and the size binning in the histogram was adapted in order to have an approximately equal number of investigated color centers per size slot. The red solid line is a fit of the data using equation~(\ref{fit}).}
\label{Fig3}
\end{figure}

\indent For a set of single NV defects, the emission spectrum and the ND size were measured. Single NV$^{0}$ and NV$^{-}$ defects were then sorted according to their photoluminescence spectrum, in order to infer the proportion $\mathcal{P}_{{\rm NV}^{-}}$ of  negatively-charged NV$^{-}$ defect as a function of the ND size. The results of this experiment give a clear evidence that $\mathcal{P}_{{\rm NV}^{-}}$ drastically decreases with the ND size (see Fig.~\ref{Fig3}). Indeed, for 100-nm nanocrystals $\mathcal{P}_{{\rm NV}^{-}}$ is close to unity  but decreases to $40$\% for 20~nm  NDs. We noted that no blinking was observed in the photoluminescence intensity for all studied NV defects. In addition, all the measurements were performed while exciting NV defects below the saturation, in order to avoid photo-ionization effects~\cite{Manson_DRM2005,Gaebel_JPhys2006}. \\

\begin {figure}[t] 
\centerline
{\includegraphics[width=8.5cm]{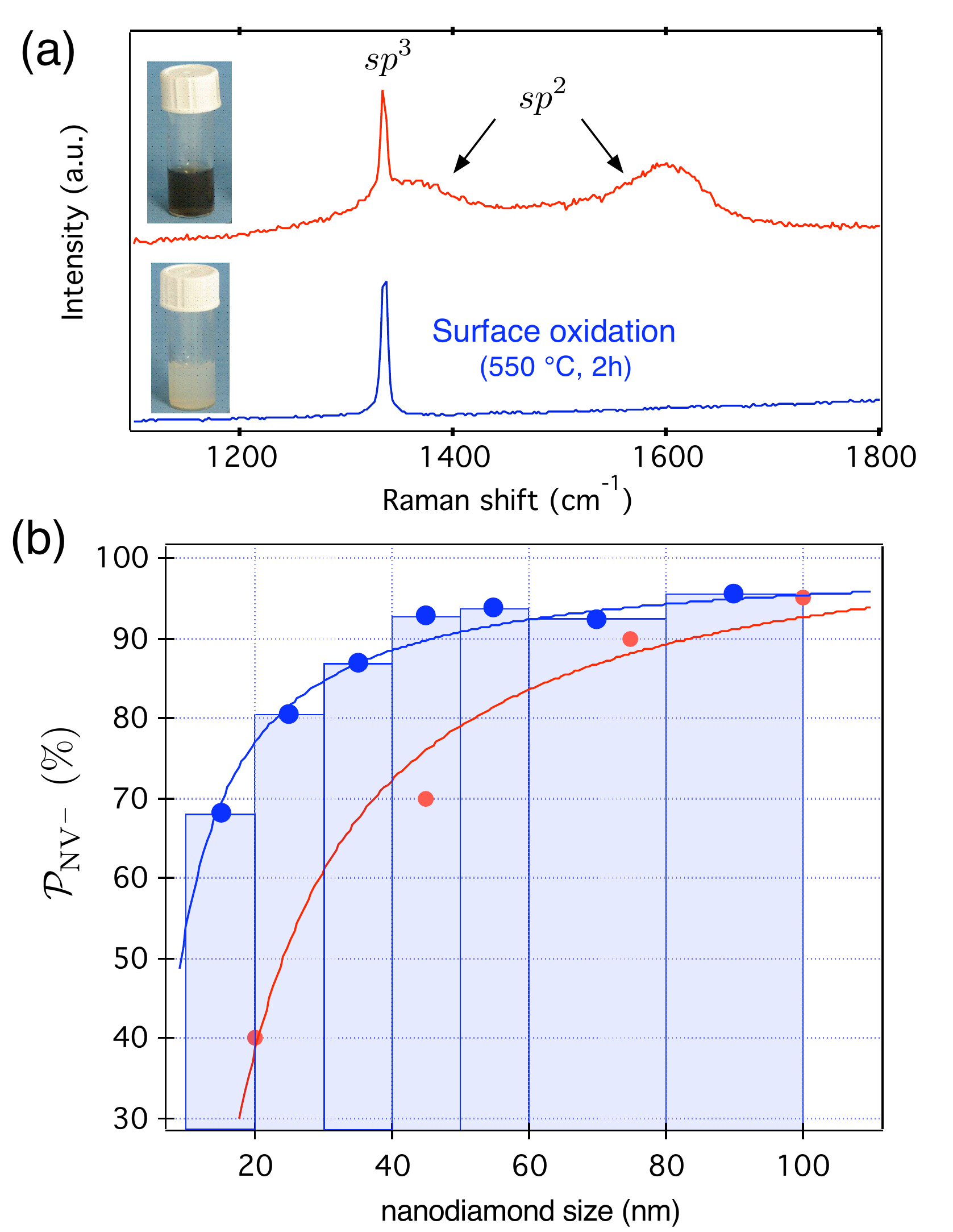}}
\caption{(color online) (a) Raman spectra associated with NDs cleaned with a piranha solution after electron irradiation and annealing (upper red curve) and with subsequent air oxidation at $550^{\circ}$C during one hour (bottom blue curve). The G-band at $1600$ cm$^{-1}$ and the D-band between $1250$ cm$^{-1}$ and $1450$ cm$^{-1}$ are associated to $sp^{2}$-hybridized carbon atoms. Air oxidation leads to a higher purity of $sp^{3}$-bonded carbon since both the D-band and the G-band are suppressed. Insets show the pictures of the aqueous colloidal solutions before and after air oxidation. (b)-Proportion of negatively-charged NV$^{-}$ defects $\mathcal{P}_{{\rm NV}^{-}}$ as a function of the ND size after air oxidation (blue points). The statistics is built from a set of $183$ single NV centers. The blue solid line is a fit obtained using equation~(\ref{fit}). The red points are reproduced from Fig.\ref{Fig3} for comparison.}
\label{Fig4}
\end{figure}

In the following, we introduce a simple model in order to explain the experimental observations. After the electron irradiation, the vacancies migrate upon annealing at $800^{\circ}$C and can be trapped by a substitutional nitrogen atom, thus forming a neutral NV$^{0}$ defect according to the reaction
\begin{equation}
{\rm N} + {\rm V} \rightarrow   {\rm NV}^{0} \ .
\label{Eq1}
\end{equation}
In order to convert the defect into its negatively-charged state NV$^{-}$, it is widely accepted that an additional electron donor located in the vicinity of the NV defect is required~\cite{Collins_Jphys2002,Collins_JAP2005}. Since the most abundant electron donors in type Ib diamond are substitutional nitrogen atoms, the process leading to the formation of NV$^{-}$ defect involves a second nitrogen atom and can be described by the ``chemical'' reaction~\cite{Waldermann_DRM2007}
\begin{equation}
{\rm NV}^{0} + {\rm N} \rightleftarrows   {\rm NV}^{-} + {\rm N}^{+}  
\label{Eq1}
\end{equation}
which can be characterized by an equilibrium constant $K$ defined as 
\begin{equation}
K=\frac{[{\rm NV}^{-}]_{eq} \times [{\rm N}^{+}]_{eq}}{[{\rm NV}^{0}]_{eq} \times [{\rm N}]_{eq} }   \ ,
\label{Keq}
\end{equation}
where $[i]_{eq}$ represents the volumic concentration of the chemical species $i$ inside the ND.\\
\indent In addition, we consider that extra electrons provided by the substitutional nitrogen atoms can be captured by electron traps T located at the ND surface. This antagonist effect can be due to graphitic defects associated to a $sp^{2}$-hybridized carbon layer  created at the ND surface during electron irradiation and annealing~\cite{Ristein_DRM2000}. Consequently, a part of nitrogen impurities can be ionized by this electronic depletion layer according to the reaction
\begin{equation}
 {\rm N} +  {\rm T} \rightarrow  {\rm N}^{+} + {\rm T}^{-} \ ,
\label{Eq2}
\end{equation}
which prevents the nitrogen impurities to donate the required electron allowing to convert the NV defect into its negatively-charged state, as represented by equation~(\ref{Eq1}).\\
\indent In the following, we assume that the reaction~(\ref{Eq2}) is total. Furthermore, the content of NV defects per unit volume is supposed to be much smaller than all the other chemical species. This assumption is verified in our experimental conditions since the study was limited to NDs hosting single NV defects. When the initial volumic concent of traps is such that $[{\rm T}]_{i}>[{\rm N}]$, all the nitrogen atoms are ionized and the charge state conversion characterized by reaction~(\ref{Eq1}) becomes fully inefficient. On the other hand if $[{\rm T}]_{i}<[{\rm N}]$, the content of ionized nitrogen at equilibrium $[{\rm N}^{+}]_{eq}$ is fixed by the reaction~(\ref{Eq2}), {\it i.e.} $[{\rm N}^{+}]_{eq}=[{\rm T}]_{i}$ and $[{\rm N}]_{eq}=\mathcal{C}_{0}-[{\rm T}]_{i}$, where $\mathcal{C}_{0}=[{\rm N}]_{eq}+[{\rm N}^{+}]_{eq}$ is the total nitrogen concentration.\\ 
\indent Since the electron traps are assumed to be located at the ND surface, their equivalent volumic content  can be written as $[{\rm T}]_{i}= \sigma_{\rm T}/a$, where $a$ is the characteristic size of the ND and $\sigma_{{\rm T}}$ the trap surface density. Using equation~(\ref{Keq}), the proportion of negatively-charged NV defects $\mathcal{P}_{{\rm NV}^{-}}$ is then given by
\begin{equation}
\mathcal{P}_{{\rm NV}^{-}}=\frac{[{\rm NV}^{-}]_{eq}}{[{\rm NV}^{-}]_{eq}+[{\rm NV}^{0}]_{eq}}=1-\frac{1}{1+K \left(\frac{\displaystyle a}{\displaystyle  a_{t}}-1\right)} \ ,
\label{fit}
\end{equation}
where $a_{t}=\sigma_{{\rm T}}/\mathcal{C}_{0}$ is a threshold size. If $a\gg a_{t}$, the number of nitrogen atoms is much higher than the number of electron traps, leading to $\mathcal{P}_{{\rm NV}^{-}}=1$. Conversely, in the limit where $a=a_{t}$, {\it i.e.} $\mathcal{C}_{0}=[{\rm T}]_{i}$, all the nitrogen atoms are ionized by the surface traps, thus preventing the formation of any negatively-charged NV defect in the nanocrystal  and $\mathcal{P}_{{\rm NV}^{-}}=0$. Using equation~(\ref{fit}) as a fitting function, a good agreement with experimental data points is achieved for $K=1.0\pm 0.5$ and $a_{t}=14\pm 1$~nm (see Fig.~\ref{Fig3}).  Considering a 200-ppm nitrogen content, which corresponds approximatively to $0.04$ nitrogen atoms per cubic nanometer, the result of the fit leads to a surface density of  electron traps $\sigma_{{\rm T}}\approx 0.5$~nm$^{-2}$.\\ 
\indent As already mentioned, graphitic related defects formed on the nanodiamond surface during electron irradiation and annealing are possible electron trap candidates~\cite{Ristein_DRM2000}. In order to check the presence of graphitic shells or amorphous carbon layer, the composition of the studied NDs was analysed by recording their Raman spectra. As shown in Fig.~\ref{Fig4}(a) (red line), the $sp^{2}$-hybridized carbon is evidenced through a broad D-line at $1350$ cm$^{-1}$ and a broad G-band around $1600$ cm$^{-1}$, while the diamond core gives the usual sharp Raman line at $1335$ cm$^{-1}$~\cite{Krueger_Carbon}. 
This result indicates that a significant content of $sp^{2}$ carbon was partly covering the ND surface, even after acid cleaning with the piranha solution.\\

\indent According to the previously developed model, the proportion of negatively-charged NV defect in NDs might be improved by reducing the surface density $\sigma_{{\rm T}}$ of electron traps. For that purpose, the graphitic shell which could be associated to the electronic depletion layer at the ND surface can be efficiently removed through surface oxidation. This can be achieved with a repetitive cleaning with a strong acid treatment, using a mixture of sulfuric, nitric and perchloric acids~\cite{Chang_NatNano2008,Tisler_ACSNano2009}. Another method, which does not require the use of toxic or aggressive chemicals, consists in heating NDs in air~\cite{Chang_NatNano2008,Osswald_JACS2005,Pichot_DRM2008}. This procedure leads to oxidation of $sp^{2}$-hybridized carbon (graphite and disordered carbon), thus resulting in a higher purity of $sp^{3}$-bonded carbon (diamond)~\cite{Smith_DRM2010}. In the following, we check if the proportion of negatively-charged NV$^{-}$ defect is modified after elimination of  $sp^{2}$-bonded carbon with air oxidation. For that purpose, the same colloidal solutions of NDs were oxidized in air at $550^{\circ}$C during two hours. We note that the stability of the ND aqueous suspensions remained unchanged after this surface oxidation.\\

\indent The efficiency of the surface oxidation was first analysed by recording the Raman spectrum of the oxidized NDs. As shown in Fig.~\ref{Fig4}(a) (blue curve), the spectral features related to $sp^{2}$-bonded carbon are greatly suppressed. The effect of surface oxidation was also confirmed by the color of the solution of NDs, which became whitish after air oxidation  (see insets in Fig.~\ref{Fig4}(a))~\cite{Osswald_JACS2005}. In addition, the surface oxidation led to a size reduction of the NDs~\cite{Chang_AdvMat2009}. Using dynamic light scattering, the mean value of the size distribution was found to be reduced from $35$~nm (resp. $65$ nm) to $27$~nm (resp. $53$ nm) for the solution of SYP 0.05 (resp. SYP 0.1).\\
\indent From a set of single NV defects in oxidized NDs, we inferred the proportion $\mathcal{P}_{{\rm NV}^{-}}$ of negatively-charged NV$^{-}$ defects as a function of the ND size (Fig.~\ref{Fig4}(b)). As previously observed, $\mathcal{P}_{{\rm NV}^{-}}$ decreases with the ND size. However, the proportion of ${\rm NV}^{-}$ defects found in small NDs is strongly enhanced compared to the non-oxidized ND solution. 
Indeed, $\mathcal{P}_{{\rm NV}^{-}}$ reaches $70\%$ for near 10-nm NDs, which is two times larger than the result obtained for non-oxidized NDs (see Fig.~\ref{Fig3}). 
By fitting the data with equation~(\ref{fit}), a good agreement  is found with parameters $K=1.4\pm 0.4$ and $a_{t}=6\pm 1$~nm  which corresponds to a surface density of charge traps $\sigma_{{\rm T}}\approx 0.25$ nm$^{-2}$. This value is twice smaller than the one obtained for non-oxidized NDs. According to modifications in the Raman spectra, we assume the existence of other trapping sites inside  the nanocrystal volume, being for instance associated to dislocations of the diamond matrix or to extended defects formed by vacancies during the annealing process~\cite{Santori_PRB2007,Hounsone_PRB2006}. \\
\indent Finally, we note that the photoluminescence intensity of single NV defects in oxidized NDs was improved by roughly 4-fold and the excited-state lifetime was found longer, as recently reported in the context of ensemble measurements~\cite{Smith_DRM2010}. Both effects can be understood by considering the layer of $sp^{2}$-hybridized carbon atoms as a photoluminescence quencher. However, a quantitative study of the modifications of the photoluminescence properties of the NDs  can be hardly done since both the emission intensity  and the excited-state lifetime depend on the dipole orientation of the NV defect which is randomly distributed over each ND. \\ 

To summarize, we have reported a study of surface-induced charge state conversion of single NV defects hosted in NDs. This study illustrates the influence of surface-related defects associated with 
$sp^{2}$-hybridized carbon on the charge state of the NV color center. 
A simple model led to a threshold size for NDs hosting negatively-charged NV$^{-}$ defects, which is directly related to the surface density of electron traps. 
By using a thermal oxidation allowing to remove the shell of amorphous carbon around the NDs, we demonstrated that this threshold size can be significantly reduced, thus leading to an enhancement of the proportion of negatively-charged NV$^{-}$ defect in NDs of near 10~nm size. These results are invaluable for further understanding, control and use of the unique properties of negatively-charged NV defects in diamond, from high-resolution magnetic sensing to bioimaging.\\

\indent The authors acknowledge F.~Jelezko for fruitful discussions. They gratefully thank F. Lain\'e and F.~Carrel for the use of the SAPHIR electron irradiation facility of CEA-LIST. This work was supported by the European Commission  project EQUIND (Sixth Framework Program, IST-034368) and  the NaroSci-ERA NEDQIT  project, by the Agence Nationale de la Recherche through the DIAMAG and NADIA projects, and by C'Nano \^Ile-de-France.

\end{document}